\documentclass[aps,12pt,final,notitlepage,oneside,onecolumn,nobibnotes,nofootinbib,superscriptaddress,noshowpacs,centertags]{revtex4}

\input{tcilatex}

\begin{document}

\title{Hydration of hydrophobic solutes treated by the fundamental measure approach}
\author{G.N. Chuev and V.F. Sokolov}

\begin{abstract}
We have developed a method to calculate the hydration of hydrophobic solutes
by the fundamental measure theory. This method allows us to carry out
calculations of the density profile and the hydration energy for hydrophobic
molecules. An additional benefit of the method is the possibility to
calculate interaction forces between solvated nanoparticles. On the basis of
the designed method we calculate hydration of spherical solutes of various
sizes from one angstrom up to several nanometers. We have applied method to
evaluate the free energies, the enthalpies, and the entropies of hydrated
rare gases and hydrocarbons. The obtained results are in agreement with
available experimental data and simulations.

\textbf{Keywords: Hydrophobic solvation, density functional theory,
fundamental measure model, hydration of hydrocarbons}
\end{abstract}

\maketitle

\address{Institute of Theoretical and Experimental Biophysics, Russian Academy
of Sciences, Pushchino, Moscow region, 142290 Russia}

\section{Introduction}

Hydrophobic interactions play an important role in stabilization of various
biomacromolecular complexes including nucleic acids,proteins,and
lipids,because these complexes contain a large number of nonpolar groups 
\cite{1,2}. Despite the long history of studies of hydrophobic interactions,
the theoretical treatment of the nature of hydrophobic interactions is still
incomplete. The main problem of the interactions is a complicated multiscale
character of these effects which can reveal as microscopic changes of water
structure near small hydrophobic groups, as well as conformations and
aggregation of biomacromolecules at mesoscopic scales up to several tens of
angstroms \cite{chan}. The Monte Carlo (MC) and molecular dynamics (MD)
methods are most frequently used for modelling molecular interactions in
solutions \cite{frenkel}. However, their application to the problem of
hydrophobic interactions of macromolecules demands huge computing expenses
and in some cases it is essentially limited because of the specified
multiscale character of these interactions.New methods based on a
statistical treatment have been actively developed in last decade \cite
{chan,hummer2,pratt,chand2,basilevsky}. The most suitable method for the
specified effects seems to be the density functional theory (DFT) \cite
{evans, barrat}. The basic purpose of the approach is to construct the free
energy functional of the system which depends on density distribution of
liquids particles and intermolecular interaction potentials. Within the
framework of this approach there are a lot of various models connected with
a concrete choice of the density functional \cite{Asc, Rosen, Rosen1, Tar}{\ 
}.

In our opinion, one of the most perspective DFT models for calculation of
solvation phenomena is the fundamental measure theory (FMT) \cite{Rosen,
Rosen1, Lowen, schmidt} which determines the free energy functional as the
sum of the weighted contributions dependent on geometrical characteristics
of fluid particles. It automatically results in definition of the weighted
functions which are responsible for the volume and the surface contributions
to the solvation energy. This approach is intimately related to the
scaled-particle theory (SPT) \cite{14a,15a} for the homogeneous hard-sphere
fluid, and thus one expects that in the uniform limit it should reproduce
the SPT results. The current status of this theory includes various
generalizations of the scheme to different inter-particle interactions,
binary mixtures and polydisperse systems, application to interfaces,
wetting, confined geometries, porous media, and dynamical problems as well
(see, review \cite{Lowen}). In this work we will calculate the radial
distribution functions for rare gases and hydrocarbons, and thermodynamic
parameters of their hydrophobic solvation on the basis of the modified FMT
which treats the pressure of the liquid system correctly as distinct from
the original FMT. In particular, we will determine the size dependence of
the solvation energy and calculate interaction forces between two solutes
within the limit of low concentration of the dissolved particles. The rest
of this paper is organized as follows. The theory is described in section
II, the obtained results and the concluding remarks are presented in section
III.

\section{Fundamental measure theory}

The DFT is based on the unequivocal dependence between equilibrium density
distribution $n_{eq}(\mathbf{r})$ and external potential $u_{ext}(\mathbf{r})
$ acting on a system \cite{mermin}. The free energy of the system $F[n]$ is
related with the thermodynamic ($\Omega $) and the chemical ($\mu $)
potentials: 
\begin{equation}
\tilde{\Omega}[n,u]=F[n]-\int d\mathbf{r}n(\mathbf{r})u(\mathbf{r}),
\label{var}
\end{equation}
where $u(\mathbf{r})=\mu -u_{ext}(\mathbf{r})$. The equilibrium density $%
n_{eq}(\mathbf{r})$ is determined by the minimum $\Omega \lbrack u]=\tilde{%
\Omega}[\rho _{b},u]$ for the given temperature $T$: 
\begin{equation}
{\frac{{\delta \tilde{\Omega}[n,u]}}{{\delta n(\mathbf{r})}}}\Bigg|_{n_{eq}(%
\mathbf{r})}=0,\qquad {\frac{{\delta F[n]}}{{\delta n(\mathbf{r})}}}\Bigg|%
_{n_{eq}(\mathbf{r})}=u(\mathbf{r}).  \label{equil}
\end{equation}
In turn chemical potential $\mu $ is determined from boundary conditions,
i.e., the equilibrium density should tend to the average density of
homogeneous liquid $n_{eq}(\mathbf{r\rightarrow \infty })\rightarrow \rho
_{b}$. Thus, having the information about functional $\tilde{\Omega}[n,u]$
and the method for evaluation of $n(\mathbf{r})$, we can calculate all
necessary equilibrium characteristics of the system. The density functional
theory \cite{evans} solves the problem by searching the free energy $F[n]$,
which consists of two contributions: the ideal ($F_{id}[n]$) and the excess (%
$F_{ex}[n])$ free energies: 
\begin{equation}
F[n]=F_{id}[n]+F_{ex}[n],\qquad \beta F_{id}[n]=\int n(\mathbf{r})\ln [n(%
\mathbf{r})\Lambda ^{3}-1]d\mathbf{r},
\end{equation}
where $\Lambda $ is the de Broglie wave length, and $\beta =(k_{B}T)^{-1}$
is the inverse temperature. This free energy is related with the grand
canonical functional, minimization of the functional leads to the
equilibrium density: 
\begin{equation}
n_{eq}(\mathbf{r})=\rho _{b}\exp [-\beta u_{ext}(\mathbf{r})+{\frac{{\delta F%
}_{ex}{[n]}}{{\delta n(\mathbf{r})}}}-{\frac{{\delta F}_{ex}{[n=\rho _{b}]}}{%
{\delta n(\mathbf{r})}}}].  \label{scf}
\end{equation}
Thus, if the functional ${F}_{ex}{[n]}$ is known we can calculate the
density profile $n(\mathbf{r})$ and then all required characteristics of the
solvation.

There are many ways for constructing ${F}_{ex}{[n]}$, most of them use the
data on functional derivatives $\partial F_{ex}/\partial n(\mathbf{r})$ and $%
\partial ^{2}F_{ex}/\partial n(\mathbf{r})\partial n(\mathbf{r}^{\prime })$.
For Lennard-Jones (LJ) fluids the excess free energy is decomposed into the
contribution from a reference system of hard spheres, and the free energy
due to attractive interactions 
\begin{equation}
{F}_{ex}{[n]}={F}_{hs}{[n]+F}_{att}[{n}].
\end{equation}
The attractive interactions are usually treated by the first-order or the
second-order perturbation theories \cite{67,68}. In the mean field
approximation the attraction potential $u_{att}(r)$ is considered as a
perturbation which gives the contribution to the free energy 
\begin{equation}
{F}_{att}{[n]}=\frac{1}{2}\int \int [n(\mathbf{r})-\rho _{b}]u_{att}({%
\mathbf{r-r}^{\prime }})[n({\mathbf{r}^{\prime }})-\rho _{b}]d{\mathbf{r}%
^{\prime }}d\mathbf{r}.  \label{Fattvv}
\end{equation}
Various functionals for inhomogeneous hard-sphere (HS) fluids have been
developed. One of such methods is the FMT \cite{Rosen, Rosen1, Lowen,
schmidt} in which the excess free energy is calculated by the use of
coarse-grained or smoothed densities. We note that originally this method
has been formulated for HS liquids. But later this method has been applied
to spheroids of rotation \cite{13a} and also for various repulsive and
attractive potentials \cite{13,14, 15}. In HS liquids the repulsive
contribution to the excess free energy is written as 
\begin{equation}
\beta {F}_{hs}{[n]}=\int \Phi \lbrack n_{i}(\mathbf{r})]d\mathbf{r},
\label{FHS}
\end{equation}
where variables $n_{i}(\mathbf{r})$ are determined as weights 
\begin{equation}
n_{i}(\mathbf{r})=\int d{\mathbf{r}^{\prime }}\,n(\mathbf{r}^{\prime })\
w^{(i)}(\mathbf{r}-\mathbf{r}^{\prime }),  \label{barra}
\end{equation}
of the density $n(\mathbf{r}^{\prime })$ averaged with weight factors $%
w^{(i)}(\mathbf{r}-\mathbf{r}^{\prime })$ depending on the fundamental
geometrical measures of fluid particles, such as volume, surface, etc. The
original Rosenfeld formulation \cite{Rosen, Rosen1} utilizes the following
weighting functions 
\begin{eqnarray}
w^{(3)}(r) &=&\Theta (\sigma /2-r),\quad w^{(2)}(r)=\delta (\sigma
/2-r),\quad w^{(1)}(r)=w^{(2)}(r)/(2\pi \sigma ), \\
w^{(0)}(r) &=&w^{(2)}(r)/(\pi \sigma ^{2}),\quad \mathbf{w}%
^{(v_{2})}(r)=\delta (\sigma /2-r)\frac{\mathbf{r}}{r},\quad \mathbf{w}%
^{(v_{1})}(r)=\mathbf{w}^{(v_{2})}(r)/(2\pi \sigma ).  \nonumber
\end{eqnarray}
where $\delta (r)$ and $\Theta (r)$ are the Dirac delta-function and the
Heaviside function, respectively, $\sigma $ is the diameter of a solvent
particle. These weight factors determine weighted densities $n_{0}(\mathbf{r}%
),\quad n_{v1}(\mathbf{r}),\quad n_{3}(\mathbf{r})$, and latter is the local
factor of packing.

Within the FMT framework \cite{Rosen, Rosen1, Lowen, schmidt} the function $%
\Phi \lbrack n_{i}]$ is determined through six weight densities $n_{i}(%
\mathbf{r})$: 
\begin{equation}
\Phi \lbrack n_{i}]=-n_{0}\ln (1-n_{3})+\frac{n_{1}n_{2}-\mathbf{n}_{v1}%
\mathbf{\cdot n}_{v2}}{1-n_{3}}+\frac{n_{2}^{3}-3n_{2}\mathbf{n}_{v2}\mathbf{%
\cdot n}_{v2}}{24\pi (1-n_{3})^{2}}.
\end{equation}
Using these equations and (\ref{scf}) we obtain the equilibrium density 
\begin{equation}
n_{eq}(\mathbf{r})=\rho _{b}\exp [-\beta u_{ext}(\mathbf{r})+\sum_{i}(\int {%
\frac{\delta \Phi ({\mathbf{r^{\prime }}})}{\delta n_{i}({\mathbf{r^{\prime }%
}})}}w_{i}({\mathbf{r-r^{\prime }}})d{\mathbf{r^{\prime }-}}\frac{\delta
F_{hs}[n=\rho _{b}]}{\delta n_{i}})].  \label{5}
\end{equation}
Although the original FMT and the SPT is restricted by the HS fluids, there
are no limitations for solute-solvent potential $u_{ext}(r)$. Hence we may
apply the FMT to treat the LJ solutes.The information on the weighted
densities also allows us to calculate the mean force potential $W(r)$
determining the interaction force between two solutes in an infinitely
diluted solution

\begin{equation}
W(\mathbf{r})=V_{uu}(\mathbf{r})-\sum_{i}(\int {\frac{\delta \Phi ({\mathbf{%
r^{\prime }}})}{\delta n_{i}({\mathbf{r^{\prime }}})}}w_{i}({\mathbf{%
r-r^{\prime }}})d{\mathbf{r^{\prime }-}}\frac{\delta F_{hs}[n=\rho _{b}]}{%
\delta n_{i}}),  \label{W}
\end{equation}
where $V_{uu}(\mathbf{r})$ is the direct intermolecular interaction
potential. In turn the excess part of thermodynamic potential $\Omega _{ex}$
determining the solvation energy is calculated as 
\begin{equation}
\Omega _{ex}=\Delta \mu _{ex}=F_{ex}-\sum_{i}n_{i}{\frac{{\delta F_{ex}}}{%
\delta n_{i}}}.  \label{solv}
\end{equation}
Since experiments are most commonly done at fixed pressure $p$, it is
convenient to introduce the decomposition of the hydration chemical
potential into the excess solvation entropy $\Delta S$ and the excess
solvation enthalpy $\Delta H$ achieved by the use of an isobaric temperature
derivative \cite{Karplus}, 
\begin{equation}
\Delta S=\left( \frac{\delta \Delta \mu _{ex}}{\delta T}\right) _{p},\quad
\Delta H=\Delta \mu _{ex}+T\Delta S.  \label{DeltaE}
\end{equation}

\section{Results and discussion}

\subsection{Application to bulk water}

The FMT reduces to the SPT in the limiting case of homogeneous liquid,
determining the pressure $p$ and the surface tension $\gamma _{\infty }$ of
the liquid at a planar wall \cite{2005} as 
\begin{equation}
\beta p_{hs}={\frac{\delta \Phi ({\mathbf{r}}\rightarrow \infty )}{\delta
n_{3}},\quad }\beta \gamma _{\infty }={\frac{\delta \Phi ({\mathbf{r}}%
\rightarrow \infty )}{\delta n_{2}}}.  \label{SPTpg}
\end{equation}
In the general case the pressure $p_{hs}$ is too high to describe fluids
under normal conditions, for example, it yields 8000 atm for the effective
HS diameter $\sigma =\sigma _{w}=2.77{\AA }$ corresponding to water at 25$%
^{0}$C \cite{Pierotti, CavTom}. On the other hand, formula (\ref{SPTpg})
underestimates the surface tension $\gamma _{\infty }$ with respect to the
experimental values. To exclude these drawbacks both the thermodynamic
parameters are considered as the fitting ones in the modified SPT models
(see, for example \cite{17, Tang}). Thus we are to modify the FMT to obtain
the realistic estimate for the surface tension. There are a lot of ways to
do it, we use the simplest one by exclusion small distances at $r<r_{cut}$
in integrals (\ref{Fattvv}), (\ref{FHS}), and (\ref{barra}). As a result,
the formulas for the thermodynamic parameters are also modified: 
\begin{equation}
\beta \hat{p}_{hs}={\frac{\delta \Phi (r=\infty )}{\delta n_{3}}}-{\frac{%
\delta \Phi (r=r_{cut})}{\delta n_{3}}},\quad \beta \hat{\gamma}_{\infty }={%
\frac{\delta \Phi (r=\infty )}{\delta n_{2}}}-{\frac{\delta \Phi (r=r_{cut})%
}{\delta n_{2}}}.
\end{equation}
We chose $r_{cut}$ to fit $\hat{\gamma}_{\infty }$ by the experimental value 
$\gamma _{exp}$. Below we will indicate that the modified pressure $\hat{p}%
_{hs}$ also reduces significantly by three orders and plays a minor role in
the hydration of hydrophobic solutes.

First, we employ the method to evaluate the thermodynamic parameters of bulk
water. The HS diameter for water was chosen $\sigma _{w}=2.77{\AA }$\ and
density of water was the $\rho _{b}\sigma _{w}^{3}=0.7$. Such choice of the
water diameter has been motivated by the solubility experiments of Pierotti 
\cite{Pierotti}. To estimate the attractive contribution (\ref{Fattvv}) we
have used the LJ parameters corresponding to the SPC/E model of water \cite
{SPC/E}. For the solution of the equation (\ref{5}) we used the Picard
iterative algorithm. On every $k$-step of iteration it was necessary to
calculate weight densities $n_{i}(r\mathbf{)}$, which are related with $%
n_{eq}(r\mathbf{)}$ through integrals (\ref{barra}). For improvement of
convergence we used the algorithm based on mixing of parts of the previous
and new iterations $n_{in}^{k+1}=\lambda n_{in}^{k}+(1-\lambda )n_{out}^{k}$%
, where $\lambda $ is the mixing parameter dependent on the bulk density.
The step of integration made $0.01\sigma _{w}$, and the number of points of
integration is $N=2^{12}$. These numerical parameters provide the relative
precision of density profiles up to $10^{-6}$ and for the chemical potential
up to 0.1 kcal/mole. Using relations (\ref{solv}) with the cut-off radius $%
r_{cut}$ we have evaluated the excess chemical potential depending on the
diameter of the HS solute: 
\begin{equation}
\Delta \mu _{ex}(\sigma _{u})=\Omega _{ex}(\sigma _{u})-\Omega _{ex}(\sigma
_{u}=0),
\end{equation}
where $\sigma _{u}$ is solute diameter. We used $r_{cut}=0.42\sigma _{w}$ to
fit the calculated surface tension to the experimental value $\gamma _{exp}$
= 102 cal/(mole ${\AA }^{2}$) \cite{AlejTension}. Figure 1 represents the
dependence of the excess chemical potential $\Delta \mu _{ex}(\sigma
_{u})/4\pi \sigma _{u}^{2}$ obtained by this fitting. Apparently from the
figure our calculations are very well compared with the MC results \cite
{HuChan} and the calculations received on the basis of the information
theory \cite{Bio}. The figure shows that on the site from zero up to 4 {\AA }
the curve behaves almost in the linear fashion. It indicates that at such
solute sizes the surface effects do not yield the appreciable contribution
to the excess chemical potential and the volume contribution has crucial
importance. Whereas for the particles, whose radius is more than 10 {\AA },
the contribution of the volume component decreases and the surface component
grows.\newline
We have also calculated the Tolman length $\delta $, which is the surface
thermodynamic property of the water vapor-liquid interface (the distance
between the equimolar surface and the surface of tension). For this purpose
we use the relation 
\begin{equation}
\Delta \mu _{ex}=8\pi R^{2}\gamma _{\infty }\left( \frac{R-2\delta }{%
2R-\sigma _{w}}\right) ,
\end{equation}
where $R=\sigma _{u}/2$ is the solute radius. As a result, we have obtained $%
\delta $ = 0.92 {\AA } which is agreed the MC simulations of the SPC/E water 
\cite{HuChan}. This value is a little bit more than that calculated in \cite
{17} ($\delta $ = 0.9 {\AA }). The pressure of bulk water is estimated as 
\begin{equation}
p=\hat{p}_{hs}+\frac{n_{b}^{2}}{2}\int u_{att}({\mathbf{r}})d\mathbf{r}.
\end{equation}
Table 1 lists the data on the above thermodynamic parameters obtained by the
FMT with and without cut of the integration range, as well as the data
derived from MC simulations \cite{HuChan}. Although the pressure obtained by
our procedure exceeds the simulated one by an order, it does not yield the
significant effect on the dependence $\Delta \mu _{ex}(\sigma _{u})$, since
the pressure effect is minor for water under normal conditions. These
results hold, however, only at temperatures near $25^{o}$ C, since the
vapor-liquid interfacial tension of water decreases monotonically with
temperature increasing, while the excess chemical potential of hard sphere
solutes exhibits a maximum with temperature increasing.

\subsection{HS and LJ solutes}

On the basis of the designed method we have carried out calculations for
various hydrophobic objects. We have considered the two models of
hydrophobic solvation. The first of them consists in that the solute is
modeled as hard sphere. In framework of this model we have obtained the
dependence of the excess chemical potential on the radius of HS solute.
Figure 2a shows the comparison of the results received by FMT and by Monte
Carlo simulation \cite{CavTom}. The discrepancy between two results are
practically missing. The FMT results a bit underestimate the excess chemical
potential for hard-sphere solute which radius more than 4 {\AA }. To justify
our results we have also calculated the enthalpic and entropic contribution
into the chemical potential for the HS solute. Figure 2b shows the
comparison of the results calculated by the modified FMT with MC results 
\cite{CavTom}. The main difference between the FMT and the MC results is
that the FMT as the SPT yields a monotonic decrease of entropy versus the HS
diameter, while the MC data indicate more complicated behavior. On the other
hand, the magnitude of the excess chemical potential for the HS solutes
which radius is a less 6 {\AA } received by MC a little bit exceed the FMT
results.

The LJ model consists in that interaction between solute and solvent
molecules is realized by the LJ potential 
\begin{equation}
u_{ext}(r)=4\varepsilon \lbrack \left( \frac{\sigma _{uv}}{r}\right)
^{12}-\left( \frac{\sigma _{uv}}{r}\right) ^{6}],
\end{equation}
where $\varepsilon _{uv}$ and $\sigma _{uv}\ $are constructed from the
corresponding parameters $\varepsilon _{u}$, $\sigma _{u}$, $\varepsilon
_{v} $ and $\sigma _{v}$\ by Lorentz-Berhlot mixing rules. Figure 3 shows
the two examples of the radial distribution functions (methane and neon)
calculated by the modified FMT and derived from the simulations \cite
{17,3DKov1}. The discrepancies between the first maxima of the peaks do not
exceed 10\%and the widths of the peaks are practically coincide. The reason
of this difference underlies that the water molecules are modeled as hard
spheres and the peaks become higher and in many instances a bit narrower.
Since we have modeled the solutes as spheres this model seems to be more
essential for calculation the excess chemical potential of the atomic
solutes. Besides we will apply this approximation to treat the hydration of
hydrocarbons also.

\subsection{Inert gases and hydrocarbons}

For the evidence of the efficiency of the FMT we have calculated radial
distribution functions for linear, branched, cyclic hydrocarbons (methane,
ethane, butane and et al.) and for rare gases in water. The LJ parameters
are presented in Table 2. On the basis of HS model we have calculated radial
distribution function and evaluated the excess chemical potential for the
hydrocarbons and rare gases. The calculation of the excess chemical
potential has been carried out with the use of the perturbation theory (PT)
to take into account the attractive contribution of the solute-solvent
interactions. In this case we have calculated it by using equation (\ref
{Fattvv}). The attractive solute-solvent contribution to the hydration free
energy has been estimated as 
\begin{equation}
F_{uv}=\rho _{b}\int g_{hs}(\mathbf{r})U_{att}(r)d\mathbf{r},
\end{equation}
\label{Fuv} where $g_{hs}(\mathbf{r})$ is correlation function of water
molecules around the HS solute \cite{Chyd}. For this contribution we have
used the Weeks-Chandler-Anderson decomposition for the LJ potential into the
attractive and the repulsive parts \cite{67}: 
\begin{equation}
u_{att}(r<2^{1/6}\sigma _{uv})=\allowbreak -\varepsilon ,\qquad
u_{att}(r\geq 2^{1/6}\sigma _{uv})=4\varepsilon _{uv}\left[ \left( \frac{%
\sigma _{uv}}{r}\right) ^{12}-\left( \frac{\sigma _{uv}}{r}\right) ^{6}%
\right] ,
\end{equation}
where $\sigma _{uv}$ and $\varepsilon _{uv}$ are the LJ diameter and well
depth, respectively. We denote this approximation as the LJPT model.

Note that the attractive contribution of solvent-solvent interactions energy
decreases from 15 to 1 percent with increasing the solute radius from 0.5 to
20 {\AA }. Opposite, the attractive part of solute-solvent energy increase
with increasing the solute radius. Its magnitude is about 80\% of the excess
chemical potential. Table 2 shows the calculated values of the excess
chemical potential, the enthalpies and the entropies for hydrocarbons. The
magnitude of the excess chemical potentials for small hydrocarbons (methane,
ethane and propane) is hardly different from experimental results. But the
difference of the excess chemical potentials for large solutes becomes more
significant. The main reason of it consist in that the solute-solvent
attractive contribution for large solutes are essential and it's necessary
to know the realistic value of $\varepsilon _{u}$. Since hydrocarbons are
molecular solutes, unfortunately we can't apply an unique LJ parameter $%
\varepsilon _{uv}$ correctly. In much the same way we have calculated the
excess chemical potential for rare gases (He, Ne, Ar, Kr and Xe). At first,
we have calculated the energy of cavity formation and two corrections which
take into account attractive contributions to the solute-solvent and
solvent-solvent interactions. Figure 4 shows the insignificant difference
between the experimental values of the excess chemical potential and
received by the modified FMT. The agreement between the experimental and the
calculated data is strongly quantitative. Table 2 shows that the excess
chemical potential with corrections of attractive contributions of the
solvent-solute and solvent-solvent energy depends on solute radius in the
''U'' form \cite{GraU}. The left and right parts of sign ''U'' put together
the rare gases and the hydrocarbons, respectively.

We have also applied the LJ model and utilized the Lennard-Jones potential
for solute-solvent interactions. The discrepancy between the experimental
and the calculated values of the excess chemical potential for the rare
gases and hydrocarbons is practically missing. Table 2 shows the LJ
parameters of the rare gases and hydrocarbons and the results of the
calculations. We have carried out the analysis of decomposition the excess
chemical potential for hydrocarbons and rare gases on entropic and enthalpic
parts. Using Eqn.(\ref{DeltaE}), we have derived the excess chemical
potential to two parts $\Delta H$ and $-T\Delta S$. Table 2 lists the data
on the calculated and the experimental excess chemical potential, the
enthalpies, and the entropies of hydration of the hydrocarbons from methane
to hexane, the branched hydrocarbons (2-methylpropane, 2-methylbutane, and
neopentane), and the cyclic hydrocarbons (cyclopentane and cyclohexane). The
FMT calculations by the LJPT and the LJ models are labeled as LJPT and LJ,
respectively. We note similarly \cite{levyhydro} that the obvious feature of
the entropic and enthalpic terms that its are greatly larger in absolute
value than the excess chemical potential. The hydration enthalpies are large
and auspicious, and the hydration entropies are large and unfavorable. The
entropic terms are marginally larger in absolute value than the enthalpic
terms resulting in the unfavorable but small hydration free energies of the
hydrocarbons. It is recognized that this behavior is typical of hydrophobic
hydration. Solvation of apolar compounds in most other solvents, in fact, is
usually accompanied by smaller enthalpic and entropic changes. Table 2
indicates the difference between the experimental \cite{Exphycarb} and the
calculated enthalpies and entropies. The discrepancy in enthalpy does not
exceed the 7 kcal/mol for all hydrocarbons. The calculated excess chemical
potential of methane is about of 3\% more positive than the experimental
value (2.01 kcal/mol). The discrepancy between the experimental and
calculated hydration enthalpies for normal hydrocarbons is less significant
then the similar quantity for the higher hydrocarbons. The difference in the
enthalpy and the entropy partially cancels each other resulting in a smaller
discrepancy in the excess chemical potential. The calculated and
experimental hydration free energies of the hydrocarbons are positive and
practically do not increase with solute size.

The fact that the excess chemical potential of ethane is less than methane
is well reproduced by our calculations. The modified FMT overestimates the
value of enthalpy change and too highly underestimate entropy loss for
molecules from methane up to ethane.Both effects bring in less favorable
hydration free energy of ethane. It appears, because the current model of
hydrocarbons should take into account a larger benefit in favorable
hydrocarbon-water interactions in going from methane up to ethane without
the further loss of entropy. The hydration enthalpies of the normal and
linear hydrocarbons have a tend to increase as solute size rises. The
calculations reproduce qualitatively the effect for the LJPT model. The
theory overestimates the magnitude of the hydration enthalpies. Fortunately,
the difference between calculated and the experimental results are rather
closer for the LJ model. The analogue behavior is seen for hydration
entropies.

Similarly such decomposition the excess chemical potential into entropic and
enthalpic part has been carried out for rare gases. Table 2 shows the
discrepancy of solvation free energy for all rare gases to be negligible.
The difference between the calculated and the experimental excess chemical
potential for all atoms does not exceed 0.33 kcal/mol at temperature 298 K.
The calculated solvation entropy are not so close to the experimental value,
unlike the calculations of the solvation enthalpy which less overestimate
the contribution of it to solvation free energy. The theory underestimates
the values of solvation enthalpy and overestimates the solvation entropy for
all rare gases.

The hydrophobic effect is frequently connected to characteristic temperature
dependences \cite{42mJ,44mJ}. One of the most surprising observations is the
entropy of transition convergence of nonpolar molecules from gas phase or
nonpolar solvent into water at temperature about 400 K. We have made the
calculations to show that the FMT is able to predict the temperature
convergence of entropy both qualitatively, and quantitatively correctly. The
calculations have been carried out at several temperatures along the
experimental saturation curve of water. Using (\ref{DeltaE}) we have
obtained the solvation entropy by taking the derivative of the chemical
potential along the saturation curve. Figure 5 shows the temperature
dependence of the entropy for the different solutes for two cases of
calculation. In the first case (Fig.5a) we calculated the excess chemical
potential without the fact that the diameter of water decreases with
temperature increasing. The entropies are large and negative at room
temperature for all the solutes and decrease in magnitude with increasing
temperature. The temperature dependence of entropies is approximately linear
with slopes increasing with the increasing solute size. Moreover, the
entropies converge at about 400 K to approximately zero entropy, although at
closer inspection the temperature range of the convergence region is several
10 K and the entropy is not exactly zero at convergence. In the second case,
we have taken into account the dependence solvent diameter on temperature 
\cite{45mJ}. Figure 5b shows that the point of entropy convergence has
shifted to region where temperature and entropy magnitude is about 470 K and
-2.5 cal/(mol K), correspondingly. The convergence region has become a bit
wider. It is significant that taking into account the contributions
solute-solvent interactions has changed both the point of entropy
convergence (about 500 K) and the width of the convergence region.

\subsection{The mean force potential for colloids}

We have to note the one more benefit of the FMT. The theory allows to
calculate the depletion forces between two the solutes surrounded solvent
particles. For macroscopic objects there are relations for calculation of
depletion force for large solutes, depending on the distance between them.
The solutes are located in environment of small solvent particles. Viewing
hard spheres, Bradley and Hamaker have received the relations which take
into account the pair interactions between macroparticles solvated in a
simple fluid. Following this approach, the depletion potential $W$ and the
depletion force $F$ between two macroparticles have been calculated
depending on the distance $h$ between the solutes \cite{israil}: 
\begin{equation}
F_{B}(h)=-\frac{4\epsilon \pi ^{2}\rho _{b}^{2}}{12h},\qquad W_{B}(h)=-\frac{%
4\epsilon \pi ^{2}\rho _{b}^{2}}{12h^{2}}.  \label{W_Bradley}
\end{equation}
\begin{equation}
F_{H}(h)=-\frac{4\epsilon \pi ^{2}\rho _{b}^{2}}{6}\left[ \frac{2}{s^{2}-4}+%
\frac{2}{s^{2}}+\ln \left( \frac{s^{2}-4}{s^{2}}\right) \right] ,\quad
W_{H}(h)=-\frac{128\epsilon \pi ^{2}\rho _{b}^{2}}{3Rs^{3}(s^{2}-4)},
\label{F_Hamaker}
\end{equation}
where $s=2R+h/R$. We have calculated the depletion force $F(h)$ on the FMT
basis with the use of the correlation functions obtained before. Apparently
from Fig.  6, the Hamaker and the Bradley approaches badly feature the
behavior of depletion forces near to the solute, yielding strongly
underestimated values, and not taking into account the oscillating character
of it. The Hamaker and the Bradley approaches adequately predict the
depletion force values only in asymptotic distances between the dissolved
particles. Unlike them the FMT allows us to take into account attractive
interactions between solutes a bit less than one nanometer. We have also
compared the results of MD simulations \cite{H1} and that obtained by the
FMT. Figure 6 shows that the first peak of FMT results is a bit narrower and
higher than MD simulation but localization of curves' zeros almost
coincides. The reason of narrowing of the peaks seems to be the modelling of
water molecules as hard spheres not as particles interacting via
Lennard-Jones potential.

\section{Summary}

In this work we have used the FMT for the quantitative description of the
hydrophobic phenomena on the basis of the density functional theory. As a
result, we have received profiles of radial distribution functions for
isolated solutes in a hard-sphere fluid interacting with solute by
Lennard-Jones potential. Using the distribution function profile, we have
constructed the dependence of the excess chemical potential on the radius of
the spherical solutes. To fit the properties of bulk water, namely its
surface tension, we have modified the FMT by cutting the radius of
integration at small distances. The excess chemical potential has been
calculated for several systems, i.e., when interaction between the solute
and solvent is simulated as the hard sphere or as the Lennard-Jones
potential. We also have shown that the obtained distribution functions
reproduce with good precision the oscillating behavior of depletion forces
for the particles dissolved in fluid. We have applied method to evaluate the
free energies, the enthalpies, and the entropies of hydrated rare gases and
hydrocarbons. The obtained results are in agreement with available
experimental data and simulations. We conclude that the original FMT which
rigorously should be applied to liquids where all the interactions are
described by HS repulsive potentials is also well relevant for realistic
water-solute potentials. However, such success requires the hard sphere
water diameter to be used as an adjustable parameter. The recent extensions
of the FMT to soft potentials indicate that the soft FMT is capable to
predict the solvent structure for soft repulsive and attractive interaction
potentials \cite{sc1,sc}. In this case the FMT yields a systematic way to
generalize the treatment for hard bodies to soft interactions. Another
bottleneck of the current implementation is the spherical shape of solutes,
since the realistic applications should treat the three-dimensional solute
structure. There are no restrictions to generalize the above approach to the
three-dimensional case, the examples of such generalization are presented in 
\cite{f,f2,f3}, however such calculations demand special algorithms and more
sophisticated computations. Thus, we think that the FMT can provide a
promising basis for the accurate study of hydrophobic molecular solutes.

\textbf{Acknowledgments.} The authors are thankful to Maxim Fedorov and
Michail Basilevsky for fruitful discussions. This work was supported by the
Russian Foundation of Basic Research.

\newpage Table 1. Thermodynamic parameters of bulk water obtained by the
simulations and by the FMT.

\begin{tabular}{|c|c|c|c|}
\hline
& MC \cite{HuChan} & FMT & Modified FMT \\ \hline
$\beta p\sigma _{w}^{3}$ & 5.17$\times $10$^{-4}$ & 6.05 & 4.94$\times $10$%
^{-3}$ \\ \hline
$\beta \gamma \sigma _{w}^{2}$ & 1.36 & 1.06 & 1.32 \\ \hline
$\delta /\sigma _{w}$ & 0.32 & 1.59 & 0.33 \\ \hline
\end{tabular}

\newpage Table 2. LJ parameters of rare gases and hydrocarbons \cite{GraU,
levyhydro}, the experimental \cite{Exphycarb} and calculated data on of the
excess chemical potential, the enthalpies, and the entropies of their
hydration.

\begin{tabular}{|lcc|ccc|ccc|ccc|}
\hline
&  &  & \multicolumn{3}{|c|}{$\Delta \mu _{ex}$, kcal/mol} & 
\multicolumn{3}{|c|}{$-\Delta H$, kcal/mol} & \multicolumn{3}{|c|}{$-T\Delta
S$, kcal/mol} \\ \hline
Molecule & \multicolumn{1}{|c}{$\sigma _{u}$, \AA} & \multicolumn{1}{|c|}{$%
\varepsilon _{u}/k_{B}$} & exp & \multicolumn{1}{|c}{LJPT} & 
\multicolumn{1}{|c|}{LJ} & Exp & \multicolumn{1}{|c}{LJPT} & 
\multicolumn{1}{|c|}{LJ} & Exp & \multicolumn{1}{|c}{LJPT} & 
\multicolumn{1}{|c|}{LJ} \\ \hline
Helium & \multicolumn{1}{|c}{2.63} & \multicolumn{1}{|c|}{6.03} & 2.75 & 
\multicolumn{1}{|c}{3.00} & \multicolumn{1}{|c|}{2.76} & -- & 
\multicolumn{1}{|c}{0.49} & \multicolumn{1}{|c|}{0.16} & -- & 
\multicolumn{1}{|c}{3.48} & \multicolumn{1}{|c|}{2.91} \\ \hline
Neon & \multicolumn{1}{|c}{2.79} & \multicolumn{1}{|c|}{35.7} & 2.67 & 
\multicolumn{1}{|c}{2.62} & \multicolumn{1}{|c|}{2.62} & 0.35 & 
\multicolumn{1}{|c}{1.23} & \multicolumn{1}{|c|}{0.91} & 3.02 & 
\multicolumn{1}{|c}{3.84} & \multicolumn{1}{|c|}{3.53} \\ \hline
Argon & \multicolumn{1}{|c}{3.41} & \multicolumn{1}{|c|}{125} & 2.00 & 
\multicolumn{1}{|c}{2.21} & \multicolumn{1}{|c|}{1.91} & 2.38 & 
\multicolumn{1}{|c}{3.09} & \multicolumn{1}{|c|}{2.99} & 4.38 & 
\multicolumn{1}{|c}{5.30} & \multicolumn{1}{|c|}{4.89} \\ \hline
Krypton & \multicolumn{1}{|c}{3.67} & \multicolumn{1}{|c|}{169} & 1.66 & 
\multicolumn{1}{|c}{1.98} & \multicolumn{1}{|c|}{1.54} & 3.20 & 
\multicolumn{1}{|c}{4.05} & \multicolumn{1}{|c|}{4.12} & 4.86 & 
\multicolumn{1}{|c}{6.04} & \multicolumn{1}{|c|}{5.67} \\ \hline
Xenon & \multicolumn{1}{|c}{3.96} & \multicolumn{1}{|c|}{217} & 1.33 & 
\multicolumn{1}{|c}{1.60} & \multicolumn{1}{|c|}{1.38} & 3.85 & 
\multicolumn{1}{|c}{5.22} & \multicolumn{1}{|c|}{5.40} & 5.18 & 
\multicolumn{1}{|c}{6.83} & \multicolumn{1}{|c|}{6.79} \\ \hline
Methane & \multicolumn{1}{|c}{3.70} & \multicolumn{1}{|c|}{157} & 2.01 & 
\multicolumn{1}{|c}{2.07} & \multicolumn{1}{|c|}{2.11} & 2.70 & 
\multicolumn{1}{|c}{3.97} & \multicolumn{1}{|c|}{1.70} & 4.71 & 
\multicolumn{1}{|c}{6.04} & \multicolumn{1}{|c|}{3.80} \\ \hline
Ethane & \multicolumn{1}{|c}{4.38} & \multicolumn{1}{|c|}{236} & 1.84 & 
\multicolumn{1}{|c}{1.61} & \multicolumn{1}{|c|}{1.69} & 3.90 & 
\multicolumn{1}{|c}{6.55} & \multicolumn{1}{|c|}{4.41} & 5.74 & 
\multicolumn{1}{|c}{8.20} & \multicolumn{1}{|c|}{6.10} \\ \hline
Propane & \multicolumn{1}{|c}{5.06} & \multicolumn{1}{|c|}{236} & 1.96 & 
\multicolumn{1}{|c}{1.67} & \multicolumn{1}{|c|}{2.18} & 4.50 & 
\multicolumn{1}{|c}{8.70} & \multicolumn{1}{|c|}{5.76} & 6.46 & 
\multicolumn{1}{|c}{10.50} & \multicolumn{1}{|c|}{7.94} \\ \hline
Butane & \multicolumn{1}{|c}{5.65} & \multicolumn{1}{|c|}{236} & 2.08 & 
\multicolumn{1}{|c}{1.45} & \multicolumn{1}{|c|}{2.09} & 6.00 & 
\multicolumn{1}{|c}{10.93} & \multicolumn{1}{|c|}{7.42} & 8.08 & 
\multicolumn{1}{|c}{12.64} & \multicolumn{1}{|c|}{9.51} \\ \hline
Pentane & \multicolumn{1}{|c}{6.16} & \multicolumn{1}{|c|}{236} & 2.33 & 
\multicolumn{1}{|c}{1.33} & \multicolumn{1}{|c|}{2.46} & 6.25 & 
\multicolumn{1}{|c}{13.21} & \multicolumn{1}{|c|}{8.81} & 8.58 & 
\multicolumn{1}{|c}{14.97} & \multicolumn{1}{|c|}{11.28} \\ \hline
Hexane & \multicolumn{1}{|c}{6.51} & \multicolumn{1}{|c|}{236} & 2.49 & 
\multicolumn{1}{|c}{3.28} & \multicolumn{1}{|c|}{2.55} & 7.00 & 
\multicolumn{1}{|c}{12.63} & \multicolumn{1}{|c|}{6.11} & 9.49 & 
\multicolumn{1}{|c}{16.47} & \multicolumn{1}{|c|}{8.66} \\ \hline
Isobutane & \multicolumn{1}{|c}{5.55} & \multicolumn{1}{|c|}{236} & 2.24 & 
\multicolumn{1}{|c}{1.62} & \multicolumn{1}{|c|}{2.22} & 4.95 & 
\multicolumn{1}{|c}{10.54} & \multicolumn{1}{|c|}{7.06} & 7.19 & 
\multicolumn{1}{|c}{12.41} & \multicolumn{1}{|c|}{9.29} \\ \hline
2-methylbutane & \multicolumn{1}{|c}{5.84} & \multicolumn{1}{|c|}{236} & 2.44
& \multicolumn{1}{|c}{3.33} & \multicolumn{1}{|c|}{2.67} & -- & 
\multicolumn{1}{|c}{9.89} & \multicolumn{1}{|c|}{4.69} & -- & 
\multicolumn{1}{|c}{13.55} & \multicolumn{1}{|c|}{7.36} \\ \hline
Neopentane & \multicolumn{1}{|c}{5.89} & \multicolumn{1}{|c|}{236} & 2.51 & 
\multicolumn{1}{|c}{3.35} & \multicolumn{1}{|c|}{2.11} & 6.10 & 
\multicolumn{1}{|c}{10.09} & \multicolumn{1}{|c|}{5.07} & 8.61 & 
\multicolumn{1}{|c}{13.78} & \multicolumn{1}{|c|}{7.17} \\ \hline
Cyclopentane & \multicolumn{1}{|c}{5.86} & \multicolumn{1}{|c|}{236} & 1.21
& \multicolumn{1}{|c}{1.40} & \multicolumn{1}{|c|}{2.03} & -- & 
\multicolumn{1}{|c}{11.82} & \multicolumn{1}{|c|}{8.06} & -- & 
\multicolumn{1}{|c}{13.55} & \multicolumn{1}{|c|}{10.09} \\ \hline
Cyclohexane & \multicolumn{1}{|c}{6.18} & \multicolumn{1}{|c|}{236} & 1.25 & 
\multicolumn{1}{|c}{1.26} & \multicolumn{1}{|c|}{2.01} & 7.45 & 
\multicolumn{1}{|c}{13.29} & \multicolumn{1}{|c|}{9.10} & 8.70 & 
\multicolumn{1}{|c}{14.97} & \multicolumn{1}{|c|}{11.12} \\ \hline
\end{tabular}

\newpage Figure captions.

Fig. 1. Dependence of the excess chemical potential/surface area of the
solute on its HS radius. The solid line is plotted for the modified FMT, the
dashed and dash-dotted lines for the modified SPC \cite{17} and the IT \cite
{Bio}, respectively, the triangles correspond to MC simulations (SPC/E
water) \cite{HuChan}.

Fig. 2. The excess chemical potential, the entropy, and the enthalpy
calculated by the MC \cite{CavTom} and by the modified FMT: a) the excess
chemical potential (solid and dashed lines plotted for the FMT and the MC
results, respectively), b) entropies (black circles) and enthalpies (white
circles). Symbols with lines and without of them denote the FMT and the MC
results, respectively.

Fig. 3. Solute-proximal water oxygen radial distribution functions for neon
and methane. The solid line corresponds the modified FMT, while the dashed
one to the MD results for Ne \cite{3DKov1} and the MC data for Me \cite{17},
respectively.

Fig. 4. The calculated and the experimental \cite{ExpRare}(solid and dashed
lines, respectively) data on the excess chemical potentials (triangles), the
enthalpies (black circles) and the entropies (white circles) for the rare
gases.

Fig. 5. Hydration entropy of HS solutes with radii corresponding
hydrocarbons (see legend) as a function of temperature along the saturation
curve of water in the case when the water diameter is independent on
temperature (a), and temperature dependent (b) like as in \cite{45mJ}.

Fig. 6. Dependence of the depletion force on the distance between two
solutes with radii $R_{u}=8.35$ {\AA } and $\beta \varepsilon _{uv}=5$, $%
\beta \varepsilon _{v}$=1, $\sigma _{v}=3.2$ {\AA }. The solid line
corresponds to the FMT results, the dashed one to the MD simulations \cite
{H1}, while triangles and circles to the calculations by the Hamaker and the
Bradley formulas.

\end{document}